\documentclass[aps,amsfonts,amssymb,superscriptaddress]{revtex4}
\usepackage{graphicx}
\usepackage{color}

\begin{document}

\title{Temporal correlations of superconductivity above the transition temperature in La$_{2-x}$Sr$_x$CuO$_4$ probed by terahertz spectroscopy}



\author{L.S. Bilbro}
\affiliation{The Institute for Quantum Matter, Department of Physics and Astronomy, The Johns Hopkins University, Baltimore, MD 21218 USA.}

\author{R. Vald\'es Aguilar}
\affiliation{The Institute for Quantum Matter, Department of Physics and Astronomy, The Johns Hopkins University, Baltimore, MD 21218 USA.}

\author{G. Logvenov}
\affiliation{Brookhaven National Laboratory, Upton, NY 11973, USA.}

\author{O. Pelleg}
\affiliation{Brookhaven National Laboratory, Upton, NY 11973, USA.}

\author{I. Bozovic}
\affiliation{Brookhaven National Laboratory, Upton, NY 11973, USA.}

 \author{N.P. Armitage}
 \affiliation{The Institute for Quantum Matter, Department of Physics and Astronomy, The Johns Hopkins University, Baltimore, MD 21218 USA.}
 
 \date{\today}


\maketitle

\textbf{The nature of the underdoped pseudogap regime of the high-temperature superconductors has been a matter of long-term debate \cite{Timusk99a,Damascelli03a,Norman05a}.  On quite general grounds, one expects that due to their low superfluid densities and short correlation lengths, superconducting fluctuations will be significant for transport and thermodynamic properties in this part of the phase diagram\cite{Uemura91a,Emery95a}.  Although there is ample experimental evidence for such correlations, there has been disagreement about how high in temperature they may persist, their role in the phenomenology of the pseudogap, and their significance for understanding high-temperature superconductivity \cite{Wang06a,Wang05a,Lee06a,CyrChoiniere09a,Li10a}.   In this work we use THz time-domain spectroscopy (TTDS) to probe the temporal fluctuations of superconductivity above the critical temperature ($T_c$) in La$_{2-x}$Sr$_x$CuO$_4$ thin films over a doping range that spans almost the entire superconducting dome ($x=0.09$ to $0.25$).  Signatures of the fluctuations persist in the conductivity in a comparatively narrow temperature range, at most 16 K above $T_c$.  Our measurements show that superconducting correlations do not make an appreciable contribution to the charge transport anomalies of the pseudogap in LSCO at temperatures well above $T_c$.}  

\bigskip

In general, continuous phase transitions are typified by fluctuations with correlation length and time scales that diverge near $T_c$. Dynamical measurements like TTDS are a sensitive probe of the onset of superconductivity \cite{Corson99a} and measure its temporal correlations on the time scales of interest.  In the presence of superconducting vortices such high-frequency measurements are not affected by effects like vortex pinning, creep, and edge barriers that often complicate interpretation of low frequency and DC results. In this study, we investigate the fluctuation superconductivity in thin films of LSCO grown by molecular beam epitaxy (MBE). This synthesis technique provides exquisite control of the thickness and chemical composition of the films; the intrinsic chemical tunability of LSCO allows us to investigate essentially the entire phase diagram.   For details on the films, see the `Supplementary  Information' (SI).

\begin{figure}[t]
\includegraphics[width=1\columnwidth]{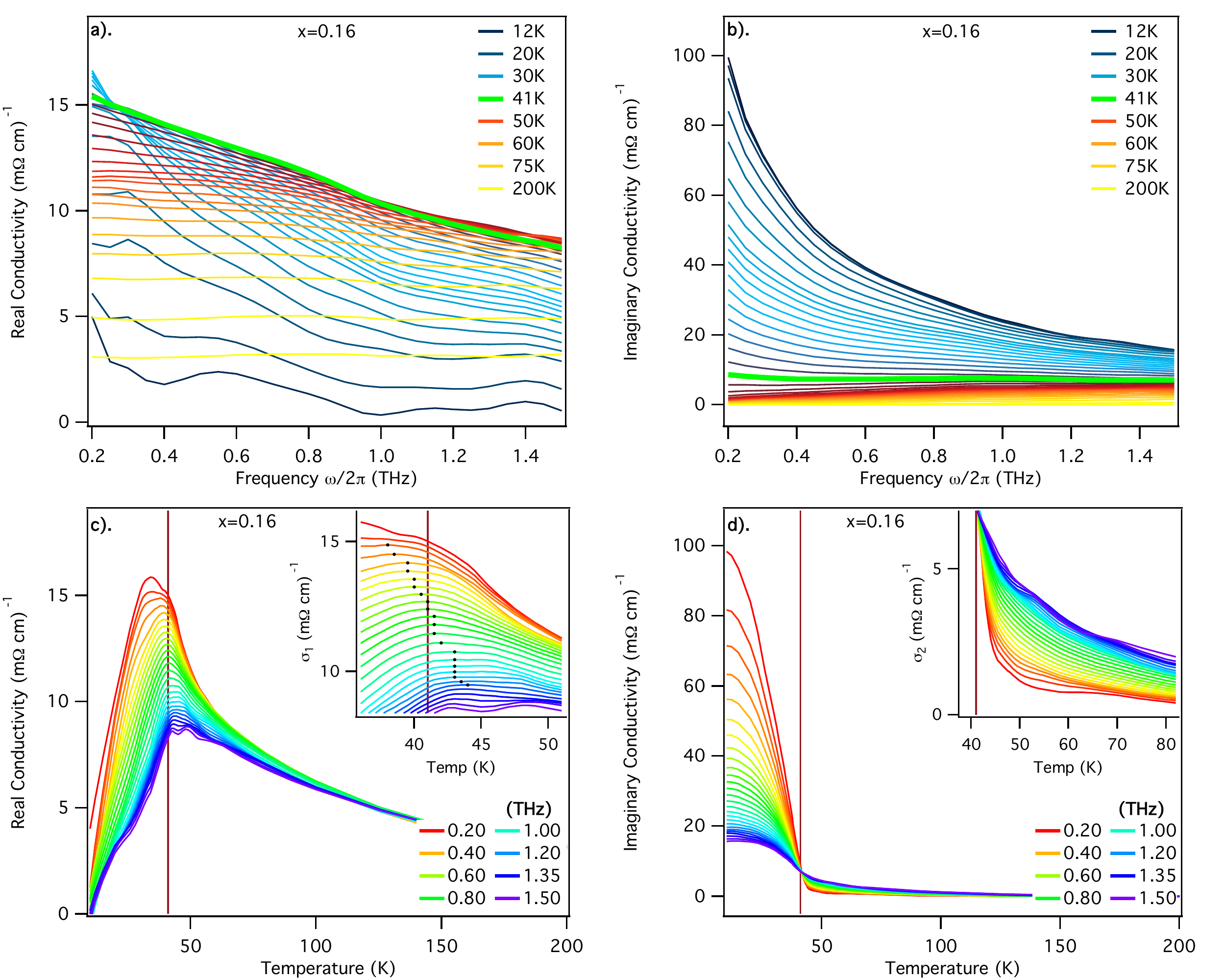}
\caption{\textsf{\textbf{THz conductivity of an $x=0.16$ La$_{2-x}$Sr$_x$CuO$_4$ thin film.}  (a), (b) Real and imaginary conductivities as a function of frequency at different temperatures for an optimally doped $x=0.16$ sample ($T_c$=41 K).  (c), (d)  Real and imaginary conductivities as a function of temperature at different frequencies for the $x=0.16$.  In panels (a) and (b) the green curve denotes $T_c$.  In panels (c) and (d) the vertical lines represent $T_c$.  Insets to (c) and (d) show expanded views of the fluctuation region.  Black dots mark the approximate peak value in the inset to (c).}}
\label{Data1}
\end{figure}

In Figs.~\ref{Data1}a and b, we show the real ($\sigma_1$) and imaginary ($\sigma_2$)  parts of the THz conductivity measured at a number of different temperatures for $optimally$ doped LSCO ($x=0.16$) with $T_c$=41 K.   We obtain similar data at other doping levels.  The spectra are easily understood in the limiting cases of high and low temperatures.  Well above the onset of superconductivity, the real part of the conductivity is almost frequency independent and the imaginary part is small, consistent with the expectation for the behavior of a metal at frequencies well below the normal state scattering rate.  At the lowest temperature the conductivity is consistent with that expected for a long-range ordered superconductor;  $\sigma_1$ is small as most of the low frequency spectral weight has condensed into the $\omega = 0$ delta function, and the frequency dependence of $\sigma_2$ is very close to $1/\omega$.   Our principal interest, however, is in the interesting transition region around $T_c$, where fluctuations of superconductivity are apparent.  Here, both components of the conductivity are enhanced, with $\sigma_1$ first rising and then falling as spectral weight moves to frequencies below the measurement range in the superconducting state.   

The fluctuation regime is perhaps most evident in Figs.~\ref{Data1}c and d, where we plot the complex conductivity of the same sample as a function of temperature.   Above the transition, the real conductivity shows a slow decrease as the temperature is raised, consistent with the increasing normal state DC resistivity.   Near $T_c$ a `loss peak' occurs due to the onset of strong superconducting fluctuations; it is exhibited at progressively lower temperatures as the probing frequency is decreased (inset to Fig.~\ref{Data1}c).  As discussed below, this is a direct consequence of the slowing down of superconducting fluctuations as $T$ is lowered.  About 10 K above $T_c$, the imaginary conductivity shows a sharp upturn at low frequency (inset to Fig.~\ref{Data1}d) indicating the onset of strong superconducting correlations \cite{Corson99a,Crane07a}.  Whether such correlations persist well above the 10-15 K range above T$_c$ is a more subtle issue, which we discuss below.  For all dopings, the region of obvious enhancement is far below the temperature of the diamagnetism and Nernst onset measured in LSCO crystals \cite{Xu00a,Wang05a,Li10a}, but is consistent with the onset found in other AC conductivity studies  \cite{Corson99a, Kitano06a,Grbic09a,Grbic10a,Maeda10a}.

An important quantity for understanding superconducting fluctuations is the phase stiffness, which is the energy scale for introducing twists in the phase $\phi$ of the complex superconducting order parameter $\Delta e^{i \phi(r)}$.   As in any continuous elastic medium one can write the energy of a phase deformation in the form of $E = \frac{1}{2} \mathcal T_\phi (\nabla \phi) ^2$ where $\mathcal T_\phi $ is a stiffness constant.   Since a phase gradient is associated with a superfluid velocity, this `elastic' energy is equivalent to the center of mass kinetic energy.   For $N$ particles of mass $m$, the stiffness is $\mathcal T_\phi =  \frac{N \hbar^2}{m}$.   One can measure this quantity directly through the imaginary part of the fluctuation conductivity $\sigma_{2f}$, as $ k_B T_\phi = \frac{\hbar \omega  \sigma_{2f} t}{G_Q} $.  Here $T_\phi$ is the 2D stiffness of a single CuO$_2$ plane given in units of degrees Kelvin, $G_Q = e^2/ \hbar$ is the quantum of conductance, and $t$ is the inter-CuO$_2$ plane spacing.  The phase stiffness is usually regarded as an equilibrium quantity;  the effect of measuring it dynamically at finite frequency $\omega$ is to introduce a length scale over which the system is probed.  In models with vortex proliferation this is typically proportional to the vortex diffusion length $\sqrt{D/ \omega}$ where $D$ is a vortex diffusion constant.

\begin{figure}[t]
\includegraphics[width=1\columnwidth]{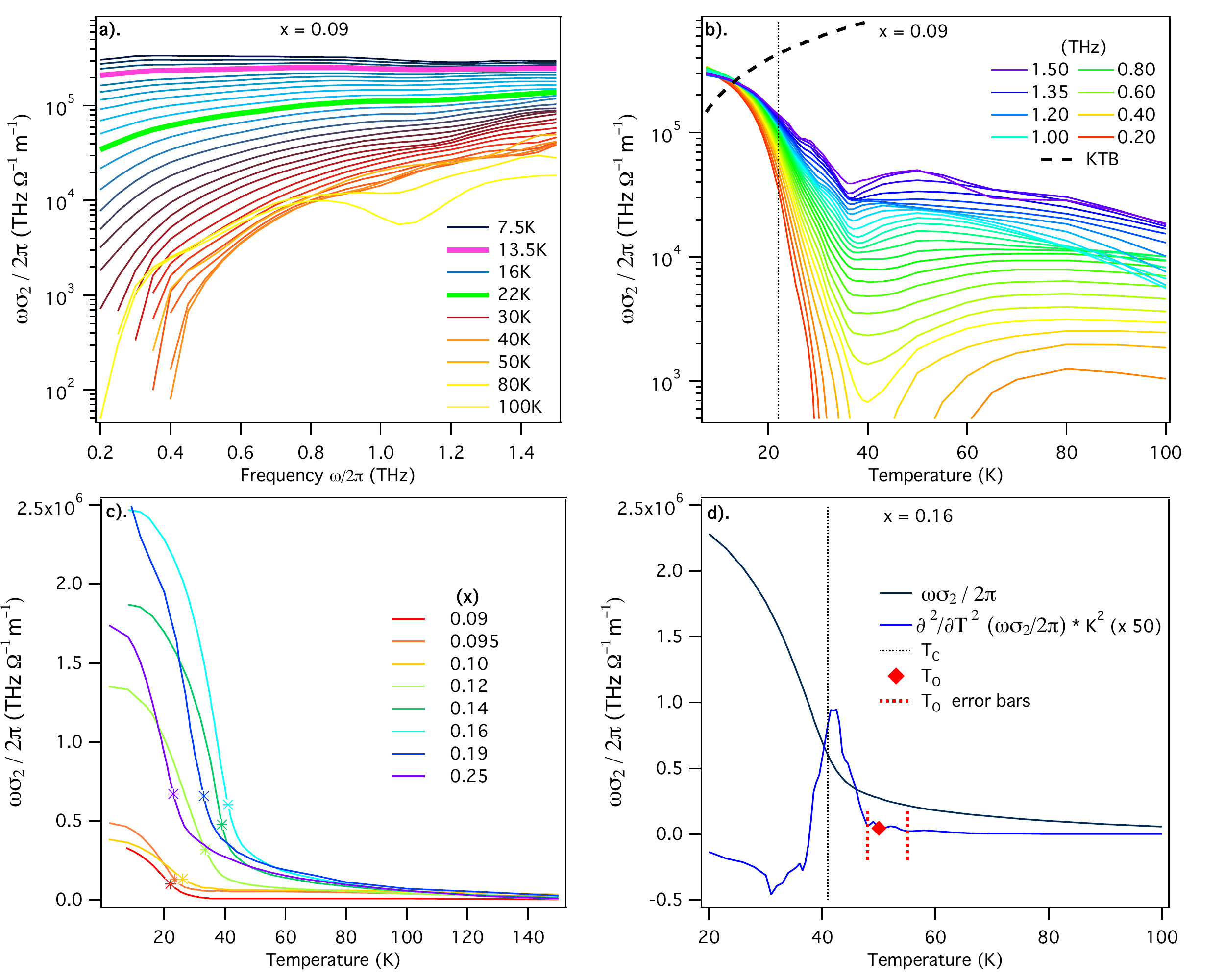}
\caption{\textsf{\textbf{$\omega\sigma_2$ is proportional to the THz `phase stiffness'} (a) $\omega\sigma_2$ vs.~frequency at different temperatures for the $x=0.09$ underdoped sample  ($T_c$=22 K).  The green curve denotes $T_c$.  The pink curve is the effective KTB temperature. (b)  $\omega\sigma_2$ as a function of temperature for different frequencies for the $x=0.09$ underdoped sample.   The dotted black line is the prediction of the KTB transition for a single isolated CuO$_2$ plane.  The temperature of its crossing with the experimental data define an effective KTB temperature.  The vertical line represents $T_c$.  (c)  $\omega\sigma_2$ at 800 GHz at a number of different dopings.   $T_c$ is indicated by the asterix and $T_o$ is indicated by diamonds.  (d) $T_o$ (red diamond) and error bars (red lines), as defined for the $x=0.16$ sample by the onset of curvature of $\omega\sigma_2$.}}
\label{Data2}
\end{figure}

In Fig.~\ref{Data2}, we plot the quantity $\omega\sigma_2$ as a function of frequency and temperature, (a) and (b) respectively, for an $underdoped$ sample ($x=0.09$) with $T_c$=22 K.  This quantity is proportional to the phase stiffness $T_\phi$ when the superconducting signal dominates over the normal state background.  Due to uncertainties in its possible form, we have abstained from subtracting a background from the plotted signal.    We find, however, that different choices for backgrounds have little affect on our conclusions.  Deep in the superconducting state, there is essentially no frequency dependence to  $\omega\sigma_2$, which is consistent with the fact that the system's phase is `stiff' on all length scales.   At higher temperatures the curves in Fig \ref{Data2}b spread as a frequency dependence is acquired.  Interestingly, the temperature where the spreading first becomes significant is very close to the temperature where the Kosterlitz-Thouless-Berenzinskii (KTB) theory (black dashed line) would predict a discontinuous jump in this quantity for an isolated CuO$_2$ plane \cite{Halperin79a}.  LSCO is of course only a quasi-2D system, but the fact that we observe a spreading at the KTB prediction shows that at some temperature $T_{KTB}^{eff} < T_c$ there begin to be significant fluctuations of a 2D character even below the transition.  $T_c$ itself does not occur until higher temperatures, as it is controlled by 3D couplings.  All of our underdoped samples, as well as previous measurements of BSCCO films \cite{Corson99a} show this behavior.  As evidenced by the frequency dependence of $\omega\sigma_2$ above $T_{KTB}^{eff}$, the fluctuations first degrade the stiffness on long length and time scales, i.e. low frequencies.  To compare across the phase diagram, in Fig.~\ref{Data2}c we plot $\omega\sigma_2$ for many dopings at a frequency 0.8 THz.  Qualitatively, we regard the onset temperature $T_o$ as the temperature where the quantity $\omega\sigma_2$ presents a substantial deviation from the trend of the higher-temperature normal state.  As a quantitative measure, we find that all samples exhibit a very sharp deviation out of the smooth high-temperature signal in this range in plots of the second derivative (Fig.~\ref{Data2}d) vs. temperature.  We take this deviation as the quantitative measure of $T_o$ (with error bars).  For all dopings, $T_o$ is no more than 16 K above $T_c$.  It is also interesting to note that the difference between $T_o$ and $T_c$ is relatively constant over the experimental range.

\begin{figure}[t]
\includegraphics[width=1\columnwidth]{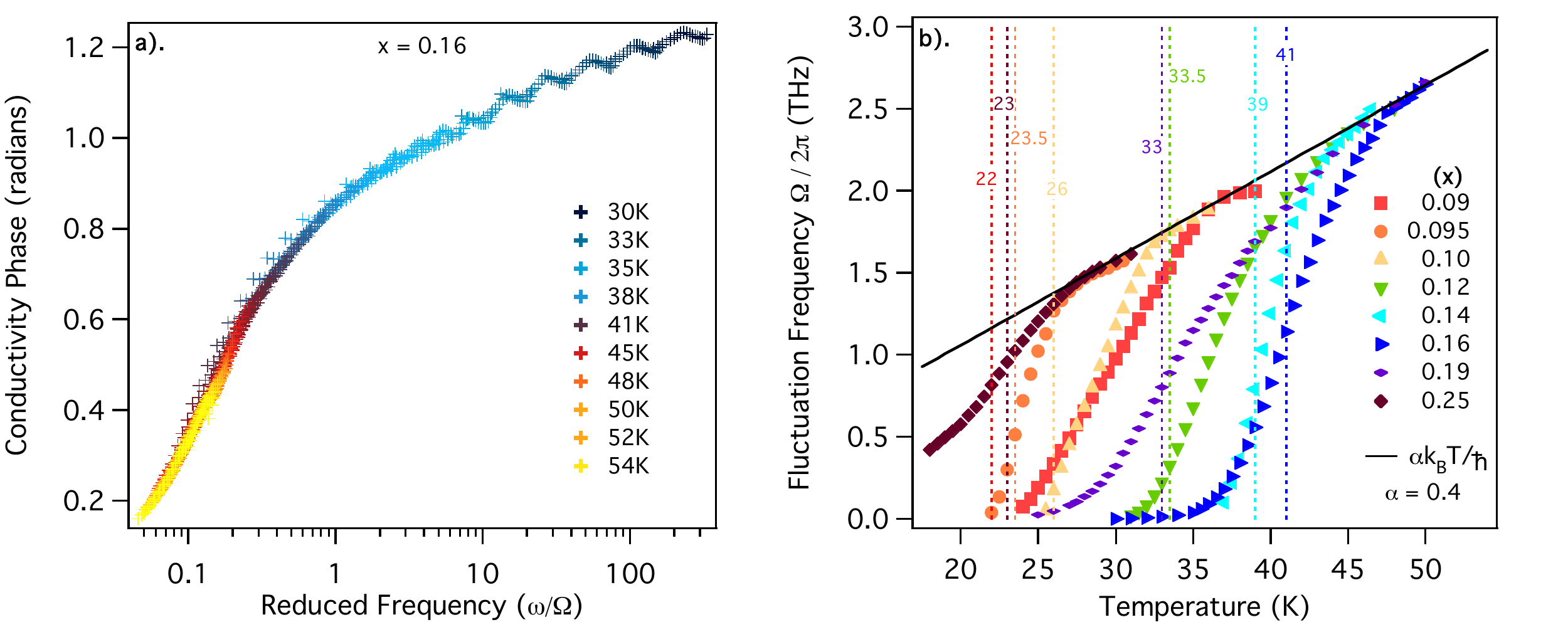}
\caption{\textsf{\textbf{Characteristic fluctuation frequency obtained through a scaling analysis.}  (a) Phase of the conductivity $\varphi =$ tan$^{-1}\sigma_2/\sigma_1$ in the frequency range from 0.5 THz - 1.5 THz as a function of the reduced frequency for 45 different temperatures in the range of 30 K - 55 K for the $x =0.16$ sample.  (b) The superconducting fluctuation rate $\Omega(T)$ determined as detailed in the text for a variety of different dopings.  Colored vertical lines are the $T_c$'s for the respective dopings.   It is observed that the fluctuation rate for all dopings approaches a limiting linear temperature dependence. }}
\label{NatureFig3}
\end{figure}

As mentioned above, the data in Figs. \ref{Data1} and \ref{Data2} are consistent with superconducting correlations that are typified by a slowing down of the characteristic fluctuation rates as temperature is decreased.   In general, the diverging length and time scales near a continuous phase transition lend themselves to a scaling analysis in which response functions can be written in terms of these diverging scales.   Close to a superconducting transition one expects that the relation 

\begin{equation}
\sigma_f(\omega)= \frac{G_Q}{t} \frac{k_B T_\phi^0}{\hbar \Omega} \mathcal S (\frac{\omega}{\Omega})
\label{scaling}
\end{equation}
 
 \noindent  holds for the portion of the conductivity due to superconducting fluctuations $\sigma_f$.  Here $T_\phi^0$ is a temperature dependent prefactor and  $\Omega$ is the characteristic fluctuation rate.   Temperature dependencies enter only through the quantities $\Omega$ and $T_\phi^0$.  This scaling function is similar to the one proposed by Fisher, Fisher, and Huse (FFH) \cite{Fisher91a} and is identical to the one used in previous THz measurements on underdoped BSCCO \cite{Corson99a}.  Note that Eq. \ref{scaling} is a very general form that does not assume any particular dimensionality of the system or character (vortex, Gaussian, etc.) of the fluctuations or functional dependencies on temperature of $\Omega$ and $T_\phi^0$.  Also note that while the fluctuation conductivity $\sigma_f = |\sigma_f| e^{i \varphi}$ is a complex quantity all prefactors to the $ \mathcal S$ scaling function are real and therefore the phase of $ \mathcal S$ is equal to the phase of $\sigma_f$.  $ \mathcal S$ is expected to exhibit single parameter scaling and thus a collapse of $\varphi$ measured at different temperatures as a function of the reduced frequency $\omega/\Omega$ yields the temperature dependent $\Omega$.  In Fig.~\ref{NatureFig3}a, we show the collapsed phase $\varphi =$ tan$^{-1}\sigma_2/\sigma_1$ of the $x=0.16$ sample as a function of the reduced frequency  $\omega/\Omega$ for 45 different temperatures in the 30 K to 55 K range.   As expected, the phase is an increasing function of $\omega/\Omega$, with the metallic limit $\varphi = 0$ reached at $\omega/\Omega \rightarrow 0$ and $\varphi$ becoming large (but bounded by $\pi/2$) as  $\omega/\Omega \rightarrow \infty$.

We were able to perform the scaling analysis and obtain similarly good data collapse for all eight samples in the range $x=0.09$ to 0.25.   In Fig.~\ref{NatureFig3}b we plot their extracted fluctuation rates $\Omega$ as a function of temperature.  In almost all samples the fluctuation rate increases very quickly (almost exponentially) within 10 K of $T_c$.  Note that the characteristic fluctuation frequency $\Omega$   is actually smooth through $T_c$, in a manner quite different than expected from the FFH scaling.   And it appears to extrapolate to zero near the effective $T_{KTB}^{eff}$ where 2D-like fluctuations become strong.  At a temperature we denote as $T_Q$, the extracted fluctuation rate crosses over to a regime where it grows at a much smaller rate that is $proportional$ to temperature as $\alpha $k$_BT/ \hbar$.  As the resistivity itself is linear in $T$ (See SI), $T_Q$ is an alternative measure of the temperature scale where we cannot distinguish superconducting fluctuations from normal state transport.   Interestingly, we continue to get good scaling and data collapse as we continue the analysis for another 5-10 K above $T_Q$.  This behavior may be related to the $\omega/T$ scaling that has been seen in the normal state \cite{Molegraaf02a}.

\begin{figure}[t]
\includegraphics[width=0.5\columnwidth]{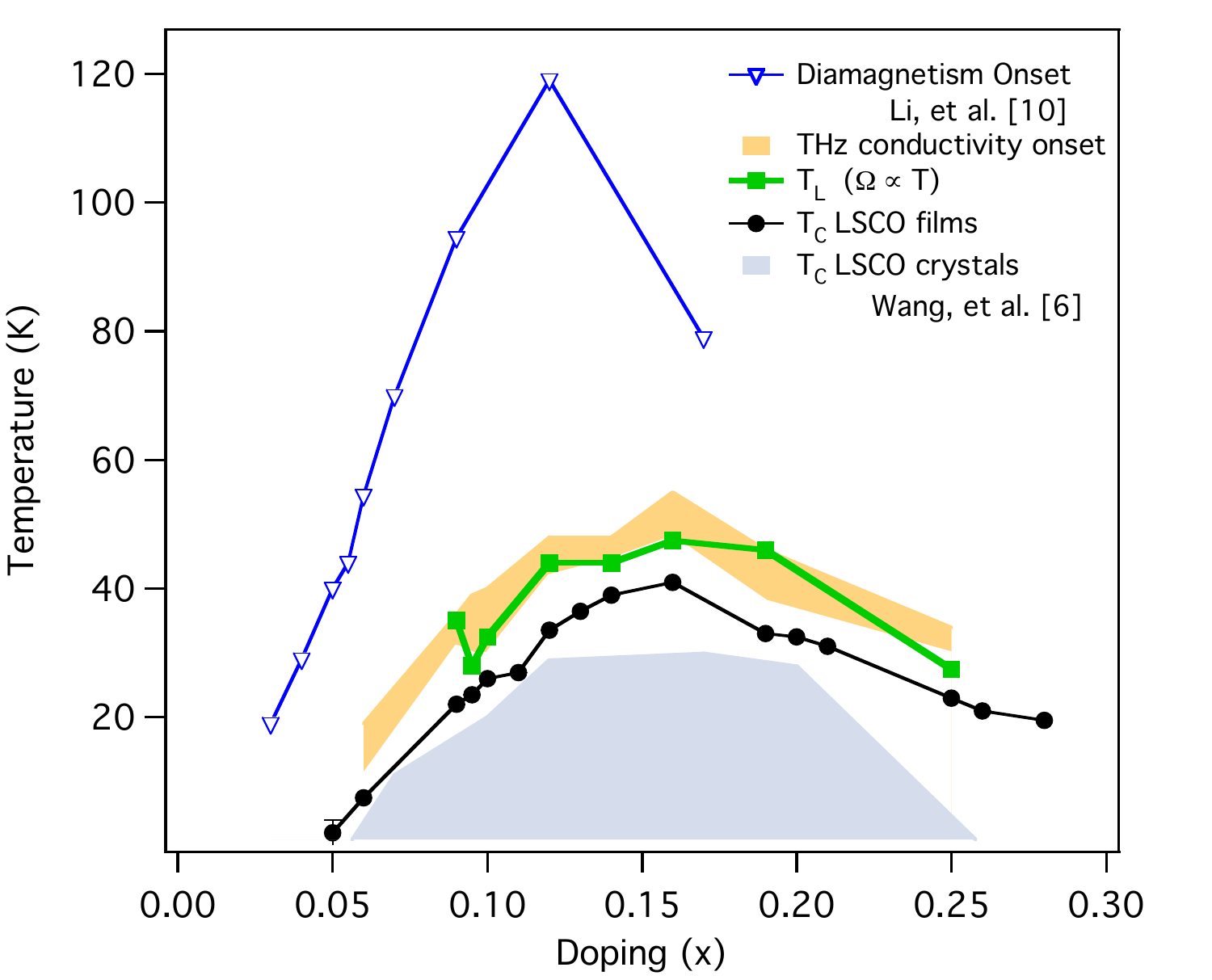}
\caption{\textsf{\textbf{Phase Diagram of the onset of superconducting correlations in LSCO.}  Temperature vs.~doping phase diagram of La$_{2-x}$Sr$_x$CuO$_4$ comparing $T_c$ of thin films and bulk crystals with the THz conductivity onset $T_o$ and the diamagnetism onset temperature from Ref. \cite{Li10a}.   Here, $T_o$ is expressed as a shaded region to convey the uncertainty in its determination.  The temperature $T_Q$ at which the characteristic fluctuation rate $\Omega$ becomes proportional to temperature is plotted in green.}}
\label{PhaseDiagram}
\end{figure}

In Fig.~\ref{PhaseDiagram} we construct a phase diagram of the fluctuation regime summarizing our results.   We compare $T_c$ of films and bulk crystals,  the diamagnetism onset from Ref. \cite{Li10a},  the region of THz conductivity onset $T_o$ (as defined by the sudden change in curvature of $\omega\sigma_2$ from Fig.~\ref{Data2}d), and $T_Q$ where $\Omega$ begins to grow linearly with temperature.  Unsurprisingly, to within experimental uncertainty $T_Q$ and $T_o$ track each other .  For all dopings the THz fluctuation conductivity shows an onset between 5 K and 16 K above $T_c$.  This is in strong contrast to measurements like diamagnetism, where the signal persists in similar samples to almost 100 K above $T_c$.  Although fluctuation diamagnetism is a thermodynamic quantity sensitive to spatial correlations and THz conductivity a dynamical quantity sensitive to temporal correlations, the differences are still surprising as within most models based on diffusive dynamics, one expects a close correspondence between them \cite{Halperin79a}.  An approximately 20 K difference in the onset temperatures exists between diamagnetism and previous THz measurements on BSCCO \cite{Li05a,Corson99a}, but the present 100 K discrepancy is much larger (particularly when normalized to the respective $T_c$'s) and cannot be easily explained away.

We may draw a few possible conclusions.   One obvious possibility is that there is little correspondence of diamagnetism (and Nernst) with conductivity because the former class of measurements are sensitive to something other than only superconductivity well above $T_c$.  In this regard, it has been shown recently that the onset of density wave order can give a strong Nernst response \cite{CyrChoiniere09a}, and that bond currents states can in principle exhibit enhanced diamagnetism \cite{Sau2010a}.  However, if the diamagnetism signal above $T_o$ is solely due to superconducting correlations then it is a well-posed theoretical challenge to explain the lack of straightforward correspondence with conductivity.  

We have found that at some temperature $T_Q > T_c$ the fluctuation rate $\Omega$ is either overwhelmed by the linear-in-$T$ normal state scattering rate or becomes linear in $T$ itself.  Calculations that model the normal state as a vortex liquid \cite{Vafek03a,Melikyan05a,Anderson08a} with a characteristic dissipation rate proportional to $T$ favor the latter scenario.  Unfortunately our measurements cannot distinguish these possibilities.     It is interesting to note that the regime in which we observe a large fluctuation conductivity is essentially the same as the regime of ``fragile London rigidity" in the diamagnetism of Li $et$ $al.$ \cite{Li05a,Li10a}.  It may be that there are two distinct types of superconducting fluctuations; one of which gives a divergent or near divergent contribution to the susceptibility in a relatively narrow range of temperatures above $T_c$ and the other of which gives a more extended, but less spectacular contribution to the susceptibility.  If the first type makes a more significant contribution to the conductivity than the second, this may resolve the apparent conflict between the two probes.  Similarly, the differences may arise in how the quantities in question depend on correlations in space (probed by diamagnetism) vs.~correlations in time (probed by conductivity).  As noted above, within classical diffusive dynamics, one generally expects a correspondence between these quantities.  Our data show that if the regime of enhanced diamagnetism $is$ due to superconductivity then a very unconventional relationship must exist between length and time correlations in the fluctuation superconductivity.   Among other possibilities, this could arise from phase separation \cite{Martin10a}, unusually fast vortices \cite{Ioffe02a,Lee03a}, or the presence of explicitly quantum diffusion.  Whatever the reason, our data show that superconducting fluctuations do not make an appreciable contribution to the charge transport anomalies in the pseudogap regime at temperatures well above $T_c$.

\section{methods}

The complex conductivity was determined by time-domain THz spectroscopy.   A femtosecond laser pulse is split along two paths and excites a pair of photoconductive `Auston'-switch antennae grown on radiation damaged silicon on sapphire wafers.    A broadband THz range pulse is emitted by one antenna, transmitted through the LSCO film, and measured at the other antenna.  By varying the length-difference of the two paths, we map out the entire electric field of the transmitted pulse as a function of time.  Comparing the Fourier transform of the transmission through LSCO to that of a reference resolves the full complex transmission.  We then invert the transmission to obtain the complex conductivity via the standard formula for thin films on a substrate:  $\tilde{T}(\omega)=\frac{1+n}{1+n+Z_0\tilde{\sigma}(\omega)d} e^{i\Phi_s}$ where $\Phi_s$ is the phase accumulated from the small difference in thickness between the sample and reference substrates and $n$ is the substrate index of refraction.  By measuring both the magnitude and phase of the transmission, this inversion to conductivity is done directly and does not require Kramers-Kronig transformation.

Our scaling analysis includes some uncertainty in setting the overall scale of $\Omega$ as Eq. \ref{scaling} only specifies $\Omega$ up to a numerical factor.  We set the overall scale of $\Omega$ so that the loss peak in Fig.~\ref{Data1}c is exhibited at a temperature when the probing frequency $\omega$ equals the fluctuation frequency $\Omega$ at that temperature.   This is an imprecise procedure and applying it strictly gives a small distribution (40$\%$) in the values of $\alpha$ for different dopings.   We choose the normalizations for the data plotted in Fig.~\ref{NatureFig3} such that  $\alpha = 0.4$ which is the mean value for all the data.   We take the relatively small spread in $\alpha$'s as evidence of the veracity of our procedure but emphasize that the $\alpha$'s could be revised with a different criteria for the scale factors.   As a crosscheck we also fit the conductivity using a Kramers-Kronig consistent fitting routine \cite{Kuzmenko05a} and get good agreement above $T_c$ between the half width of the Lorentzian peak used to model the fluctuation contribution and the rates given in Fig.~\ref{NatureFig3}b.

The LSCO films were deposited on 1-mm-thick single-crystal LaSrAlO$_4$ substrates, epitaxially polished perpendicular to the (001) direction, by atomic-layer-by-layer molecular-beam-epitaxy (ALL-MBE) \cite{Bozovic01a}.  The samples were characterized by reflection high-energy electron diffraction, atomic force microscopy, X-ray diffraction, and resistivity and magnetization measurements, all of which indicate excellent film quality. For accurate determination of the conductivity, it is critical to know the film thickness accurately. This was measured digitally by counting atomic layers and RHEED oscillations, as well as from so-called Kiessig fringes in small-angle X-ray reflectance and from finite thickness oscillations observed in XRD patterns.  The x=0.14, 0.16, and 0.25 films are 80 monolayers thick; the x=0.095 and 0.19 flims are 114 monolayers; the x=0.09, 0.10 and 0.12 films are 150 monolayers (one monolayer is $\approx 6.6 \AA$ ).




\section{addendum}
The corresponding author for this work is N.P. Armitage.

The authors would like to thank P.W. Anderson, A. Auerbach, A. Dorsey, N. Drichko, S. Kivelson, L. Li, W. Liu, V. Oganesyan, N.P. Ong, J. Orenstein, F. Ronning, O. Tchernyshyov, Z. Tesanovic, D. van der Marel, and J. Zaanen for helpful discussions and/or correspondences.   Support for the measurements at JHU was provided under the auspices of the ``Institute for Quantum Matter" DOE DE-FG02-08ER46544.   The work at BNL was supported by U.S. DOE under Project No. MA-509-MACA.

$Competing$ $Interests:$ The authors declare that they have no
competing financial interests.

$Correspondence:$ Correspondence and requests for materials
should be addressed to NPA (email:npa@pha.jhu.edu).

\section{Author Contribution}
LSB designed and built the THz spectrometer; LSB and RVA performed the THz measurements; LSB analyzed the data; GL and IB synthesized the films (using RHEED for in-situ characterization) and measured mutual inductance; OP took the XRD and AFM data; LSB, RVA, IB, and NPA wrote and revised the manuscript; NPA devised the experiment.

\newpage

\section{Supplementary Information: Film characterization by XRD, AFM, mutual inductance and resistivity measurements}

\begin{figure}[h]
\includegraphics[width=0.5\columnwidth]{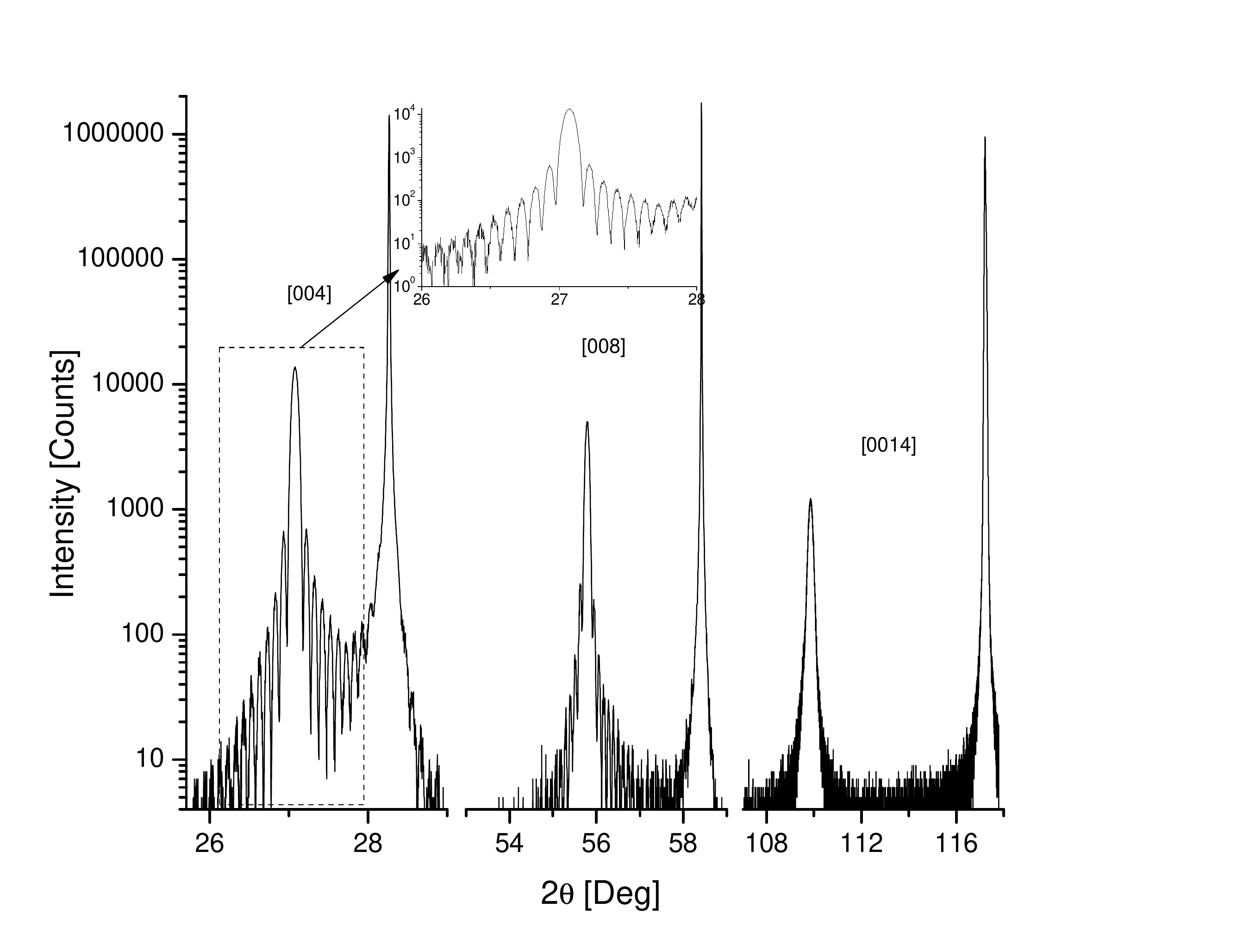}
\caption{\textsf{\textbf{X-ray diffractogram of the LSCO $x=0.1$ film.}  An accurate value of the lattice constant $c_0$ can be retrieved by means of the Nelson-Riley algorithm using a number of observed diffraction peaks. Inset: Magnified view of the (004) Bragg peak, showing pronounced finite-thickness oscillations.}}
\label{XRD}
\end{figure}

We have characterized all samples by X-ray diffraction and low-angle X-ray reflectance using an X-pert PRO PANalytical four-circle diffractometer with a Cu tube. A typical X-ray diffractogram for the $x=0.1$ sample is shown in Figure \ref{XRD}. Out-of-plane lattice constants $c_0$ are determined using multiple Bragg diffractions and applying the Nelson-Riley algorithm  (1). For the various samples, the nominal thickness is known by virtue of our digital (atomic-layer-by-layer) molecular beam epitaxy technique. This nominal thickness was in agreement to within $\pm5 \%$ with the thickness calculated from the finite-thickness oscillations (so-called Kiesig fringes) seen in the diffractograms as satellite peaks on both sides of the strong Bragg peaks (see the inset to Figure \ref{XRD}), as well as in low-angle X-ray reflectance. The error in thickness determination (typically of the order of 5-10 \AA) depends on the accuracy that one can measure the period of oscillations. The large number of well-resolved satellite peaks in Fig.~\ref{XRD} indicates that the surface roughness of the film is small.

\begin{figure}[h]
\includegraphics[width=0.5\columnwidth]{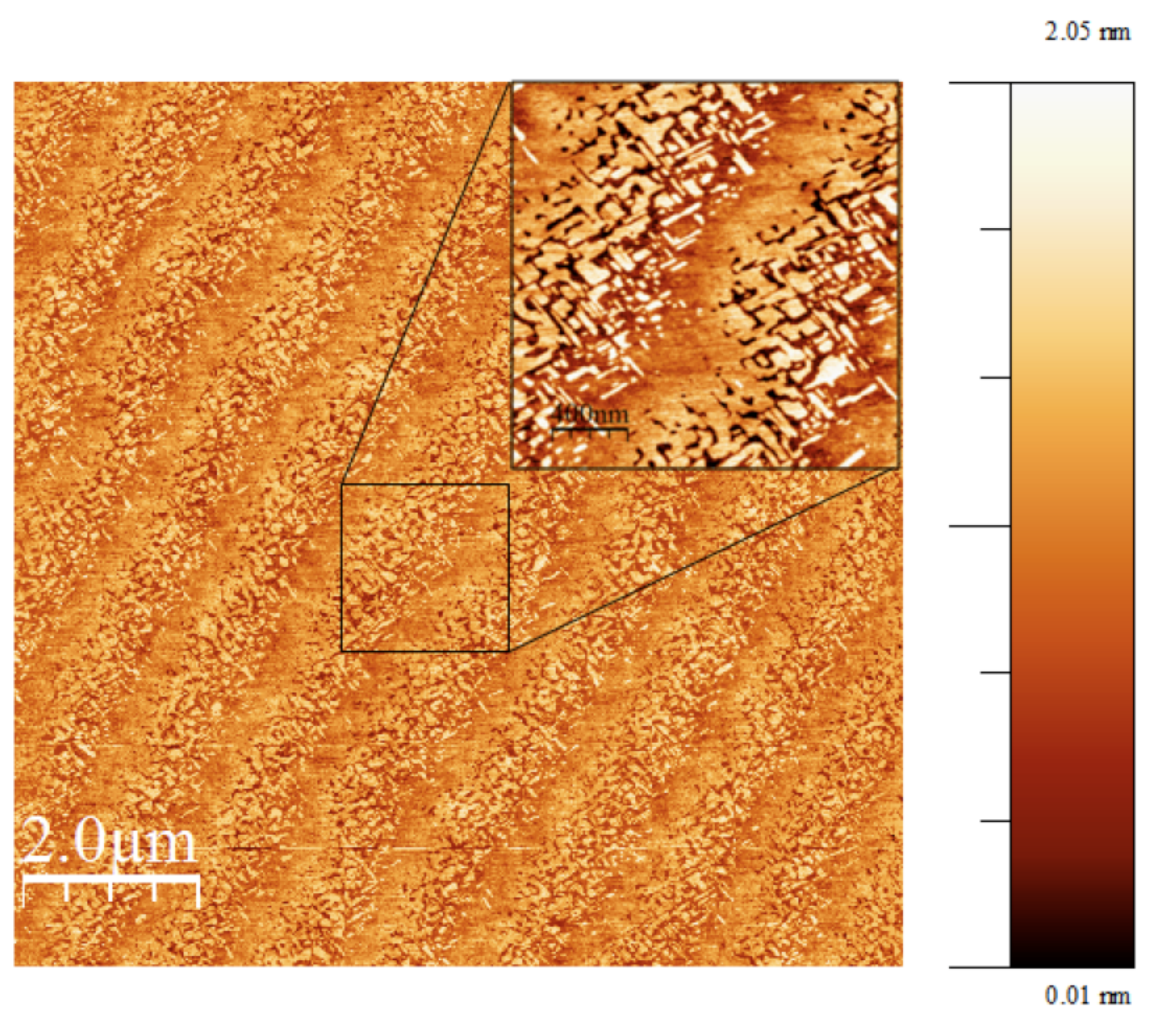}
\caption{\textsf{\textbf{Typical AFM image of the $x=0.1$ LSCO films.}  The surface of this sample show terraces that are one mono-layer (1/2 unit cell) high. The rms roughness extracted from this $10\times10$ $\mu$m image is 2.74 \AA}}
\label{AFM}
\end{figure}

This small surface roughness was confirmed with Atomic Force Microscopy (AFM). For the $10 \times 10$ $\mu$m$^2$ field of view shown in Fig.~\ref{AFM} for the $x=0.1$ sample, the typical rms surface roughness (2) varied between 0.2 and 0.9 nm; in all cases it was less than one unit cell height ($c_0 \approx$ 1.3 nm). 

\begin{figure}[h]
\includegraphics[width=0.5\columnwidth]{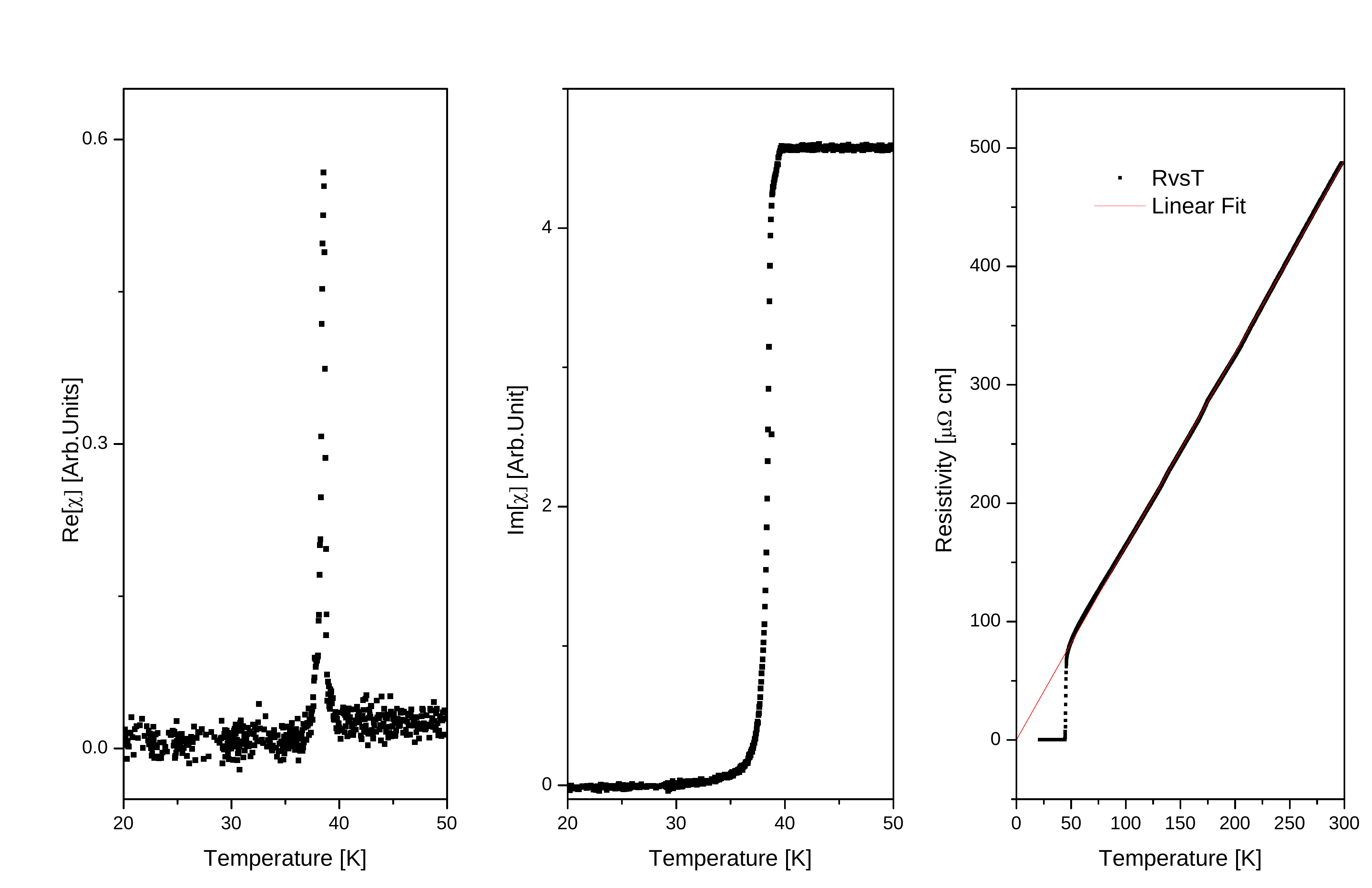}
\caption{\textsf{\textbf{Transport in LSCO thin films.} (a) The in-phase (dissipative) and (b) the out-of-phase (inductive) components of mutual inductance measured by two-coil technique. (c) Resistivity vs. temperature measured in an optimally doped LSCO film. The red straight line is the linear fit from the room temperature down to few degrees above $T_c$.}}
\label{MI}
\end{figure}

Every sample was also characterized using a two-coil transmission mutual inductance (MI) technique to check film homogeneity. The diamagnetic MI signal onsets when the resistance of the sample reaches zero. The in-phase (dissipative) MI shows a peak below $T_c$, the width of which is sensitive to both in-plane and out-of-plane inhomogeneity up to the scale of the coil diameter ($\approx$ 1 mm).  As seen in Fig.~\ref{MI}a for a optimally doped sample the measured FWHM values for the peaks were in the range between 0.2-0.5 K, which demonstrates the high homogeneity of the films and their excellent quality.  Any variation in stoichiometry, either in-plane or out-of-plane, would result in broadening of the dissipative peak, or even in the appearance of multiple peaks.

Measurements of resistance as a function of temperature provide another figure of merit attesting to the high quality of the films, the Residual Resistivity Ratio (RRR). As seen in Fig.~\ref{MI}c, for an optimally doped LSCO film the resistivity $\rho$(T) is typically linear in $T$ down to temperatures near $T_c$. This linear dependence can be extrapolated down to $T=0$, to define the so-called Residual Resistivity $\rho_0$. The ratio $\rho_0/\rho$(T=300 K) can be used as a measure of film quality.   In general a RRR of 5-10 $\%$ is considered very good. In this particular sample the RRR $\ll 1\%$.

\pagebreak

\textbf{Supplementary Information References}

\bigskip

[1] Lipson, H. \& Steeple, H \textit{Interpretation of X-ray Powder Diffraction Patterns} (Macmillan, London, 1970).

[2] Horcas, I. et al. Wsxm: A software for scanning probe microscopy and a tool for nanotechnology. \textit{Review of Scientic Instruments }\textbf{78}, 013705 (2007).

\end{document}